\documentclass[aps,twocolumn, showpacs,superscriptaddress]{revtex4-2}
\usepackage[utf8]{inputenc}
\usepackage{graphicx,float}

\usepackage{amsmath}
\usepackage{amssymb}
\usepackage{bm}
\usepackage{bbm}
\usepackage{dsfont}
\usepackage{bbold}
\usepackage{multirow}
\usepackage{color}
\usepackage[usenames,dvipsnames]{xcolor}

\usepackage[colorlinks=true,citecolor=blue,urlcolor=blue]{hyperref}

\usepackage{hyperref}
\hypersetup{colorlinks=true,linkcolor=blue,citecolor=blue,urlcolor=blue}

\newcommand{\be}{\begin{equation}}
\newcommand{\ee}{\end{equation}}

\newcommand{\sket}[1]{{\ensuremath{\lvert#1\rangle}}}
\newcommand{\lket}[1]{{\ensuremath{\left\lvert#1\right\rangle}}}
\newcommand{\ket}[1]{\if@display\lket{#1}\else\sket{#1}\fi}

\newcommand{\sbra}[1]{{\ensuremath{\langle#1\rvert}}}
\newcommand{\lbra}[1]{{\ensuremath{\left\langle#1\right\rvert}}}
\newcommand{\bra}[1]{\if@display\lbra{#1}\else\sbra{#1}\fi}

\newcommand{\sbraket}[2]{{\ensuremath{\langle#1\rvert#2\rangle}}}
\newcommand{\lbraket}[2]{{\ensuremath{\left\langle#1\!\left\rvert\vphantom{#1}#2\right.\!\right\rangle}}}
\newcommand{\braket}[2]{\if@display\lbraket{#1}{#2}\else\sbraket{#1}{#2}\fi}

\newcommand{\sketbra}[2]{{\ensuremath{\lvert #1\rangle\!\langle #2\rvert}}}
\newcommand{\lketbra}[2]{{\ensuremath{\left\lvert #1\right\rangle\!\!\left\langle #2\right\rvert}}}
\newcommand{\ketbra}[2]{\if@display\lketbra{#1}{#2}\else\sketbra{#1}{#2}\fi}

\begin{document}


\title{Quantum random number generation based on a perovskite light emitting diode}


\date{\today}

\author{J. Argillander}
\affiliation{Department of Electrical Engineering, Link\"oping University, 581 83 Link\"oping, Sweden}

\author{A. Alarc\'{o}n}
\affiliation{Department of Electrical Engineering, Link\"oping University, 581 83 Link\"oping, Sweden}

\author{C. Bao}
\affiliation{Department of Physics, Chemistry, and Biology (IFM), Link\"oping University, Link\"oping 581 83,
Sweden}
\affiliation{National Laboratory of Solid State Microstructures, School of Physics, Nanjing University, Nanjing, 210093, China}

\author{C. Kuang}
\affiliation{Department of Physics, Chemistry, and Biology (IFM), Link\"oping University, Link\"oping 581 83,
Sweden}

\author{G. Lima}
\affiliation{Departamento de F\'{\i}sica, Universidad de Concepci\'on, 160-C Concepci\'on, Chile}
\affiliation{Millennium Institute for Research in Optics, Universidad de Concepci\'on, 160-C Concepci\'on, Chile}

\author{F. Gao}
\email{feng.gao@liu.se}
\affiliation{Department of Physics, Chemistry, and Biology (IFM), Link\"oping University, Link\"oping 581 83,
Sweden}

\author{G. B.~Xavier}
\email{guilherme.b.xavier@liu.se}
\affiliation{Department of Electrical Engineering, Link\"oping University, 581 83 Link\"oping, Sweden}


\begin{abstract}
The recent development of perovskite light emitting diodes (PeLEDs) has the potential to revolutionize the fields of optical communication and lighting devices, due to their simplicity of fabrication and outstanding optical properties. Here we demonstrate, for the first time, that PeLEDs can also be used in the field of quantum technologies by demonstrating a highly-secure quantum random number generator (QRNG). Modern QRNGs that certify their privacy are posed to replace widely adopted pseudo and true classical random number generators in applications such as encryption and gambling, and therefore, need to be cheap, fast and with integration capabilities. Using a compact metal-halide PeLED source, we generate random numbers, which are certified to be secure against an eavesdropper, following the quantum measurement-device-independent scenario. The obtained random number generation rate of more than 10 Mbit s$^{-1}$, which is already comparable to actual commercial devices, shows that PeLEDs can work as high-quality light sources for quantum information tasks, thus paving the way for future developments of quantum technologies. Lastly, we argue that the simpler PeLED manufacturing process, when comparing to solid-state devices, may have large environmental impacts when quantum technology systems become more mass produced, due to the possible lower carbon footprint.

\end{abstract}




\maketitle


\section{Introduction}
The recent development of perovskite photonic devices has had significant impacts in many fields due to their optical properties and low-cost fabrication \cite{Sutherland_2016, Zhang_2020}. Perovskites kick-started a revolution as the building material for efficient solar cells \cite{Kojima_2009}, while more recently they are also showing great promise for the fabrication of efficient light-emitting diodes (LEDs) \cite{Liu_2020}. Perovskite LEDs (PeLEDs) that operate at room temperature have already been developed \cite{Tan_2014}, opening up the path for practical perovskite-based light sources. These devices have the potential to be a game changer in fields such as next-generation displays \cite{Quan_2019}, general-purpose lighting \cite{Bidikoudi_2018} and optical communications \cite{Bao_2020}, because PeLEDs possess narrow emission width and widely tunable bandgap, further motivating their use as practical devices \cite{Zhang_2020, Quan_2019}. Another key advantage is that the fabrication process of perovskite devices can be solution-processed under ambient conditions, for example through spin-coating or inkjet printing. This allows the: (i) integration of perovskite photonics with complementary metal-oxide semiconductors without relying on epitaxial growth and opens up for emerging technologies in wearable, implantable and portable electronics; and (ii) a simpler manufacturing process, which is highly attractive due to a lower associated cost \cite{Ren2021,Vasilopoulou2021}.\textcolor{black}{Finally, recent results show advantages in the life cycle assessment (LCA) of perovskite photovoltaic cells compared with a silicon benchmark in terms of reduced energy payback time and reduced greenhouse gas emission factor \cite{Tian_2020, Tian_2021}, thus pointing out that PeLEDs might also have similar advantages in their LCA when compared with standard semiconductor solid-state devices, already pointing towards a more sustainable photonics industry.}

The field of quantum information is largely dependent on technological advances provided by other areas in engineering and natural sciences. We have known for decades that the use of quantum systems for information processing opens up previously unreachable results such as carrying out some computation tasks under practical time \cite{Shor_1994} and the demonstration of the non-local behaviour of nature \cite{Bell_1964}. However it was not until very recently that experimental demonstrations of a quantum advantage in computing \cite{Arute_2019, Zhong_2020} and conclusive non-locality tests \cite{Hensen_2015} could be demonstrated. Major advances pushing forward many new applications in photonic quantum technologies include highly efficient single-photon detectors \cite{Marsili_2013}, integrated photonic circuits \cite{Wang_2020} and novel optical fibre designs \cite{Xavier_2020}. Some preliminary efforts have been made in order to use perovskite devices for quantum optics \cite{Zhang_2021,Utzat_2019}. However, perovskite-based devices are yet to be demonstrated in a complete quantum information processing task.

In our work we employ a metal-halide PeLED as the light source for a quantum random number generator (QRNG), providing a radical departure from the usual approach where solid-state light sources are usually employed (semiconductor LEDs or lasers) in quantum technology applications. The generation of random numbers based on the outputs from measurements of quantum systems is highly attractive for practical applications, due to the unavoidably random nature of quantum processes \cite{Collantes_2017}. These numbers are used for a wide variety of applications, from encryption, on-line gaming and gambling, to computer simulations. Here we are able to produce a high-quality stream of random numbers from projective measurements on weak coherent polarisation states produced from the PeLED. The random bits are certified by the widely adopted National Institute of Standards and Technology (NIST) randomness test suite \cite{NIST}. Simultaneously, we resort to the measurement device-independent (MDI) \cite{Supic_2017} approach to certify the amount of private randomness that our system can produce in the hypothetical case that our detectors could be compromised by an eavesdropper. The MDI approach is of high relevance for QRNGs as the detectors are the devices that are most prone to being hacked by side-channel attacks \cite{Lo_2012}. Our demonstration on the use of PeLEDs for the first time in a complete quantum information task paves the way for perovskite-based quantum technologies, opening up new possibilities towards the widespread deployment of quantum applications. Furthermore, the marriage of perovskite-based photonic devices with quantum information systems sets a path forward towards a more sustainable quantum technology industry, which will arguably become an issue as all areas of society must continuously lower their carbon footprint.

\section{Quantum random number generation}

Intuitively, randomness is associated with the difficulty of characterising the evolution of all the parameters leading to a given complex process. This is the basic concept behind many commercial random number generators. Nonetheless, from a more stringent perspective of cryptography, it is desired that the privacy of the generated randomness can be certified even against an eavesdropper that has some extra knowledge of the device inner-workings and infinite computational power to analyse the data generated \cite{Acin_2016}. This is only possible within the framework of quantum mechanics while exploiting the intrinsic randomness of the results of a quantum measurement \cite{Stefanov_2000}. QRNGs are usually classified in terms of the amount of assumptions needed by the user to guarantee the privacy of the generated random bits against a malicious eavesdropper. QRNGs based on the modern device-independent (DI) approach provide the ultimate level of security, guaranteeing private randomness even against an eavesdropper that has fabricated the device itself \cite{Pironio_2010, Bierhorst_2018, Liu_2018, Liu_2021, Li_2021, Shalm_2021}. However, the requirements are quite demanding and they can only achieve modest random number generation rates. On the other end of the security spectrum lie standard device-dependent (DD) QRNGs that are capable of producing high random bit generation rates, but where full characterisation of the entire system is needed \cite{Ma_2005, Dynes_2008, Wayne_2009, Furst_2010, Nie_2014}. Solutions based on reasonable assumptions about partial knowledge of the device, the so-called semi-device-independent (SDI) approaches \cite{Pawlowski_2014}, are the new trend for randomness generation as a high output throughput can be achieved with a higher level of security compared to traditional QRNGs\textcolor{black}{\cite{Lunghi_2015, Cao_2016, Nie_2016, Brask_2017a, Li_2019, Brask_2017b, Rusca_2019, Carine_2020, Mironowicz_2021, Pivoluska_2021, Zhang_2021b, Liu_2022}.} These quantum random number generators represent an important breakthrough in physics and are strong contenders to replace the ones being used today in industry and academia. Nonetheless, this will only be possible if they are made compact, have low manufacturing cost, and can be integrated with other systems.\textcolor{black}{If they are to be mass-produced, then the environmental footprint also becomes relevant. The use of PeLEDs for building modern QRNGs are therefore of ultimate importance since they can satisfy all these requirements simultaneously. We gather the generated raw bitrate of the previously mentioned QRNGs in Fig. \ref{Fig1}, separated by security class, showing clearly that our result is already highly competitive. }

\begin{figure}[ht!]
\centering
\includegraphics[width=\linewidth]{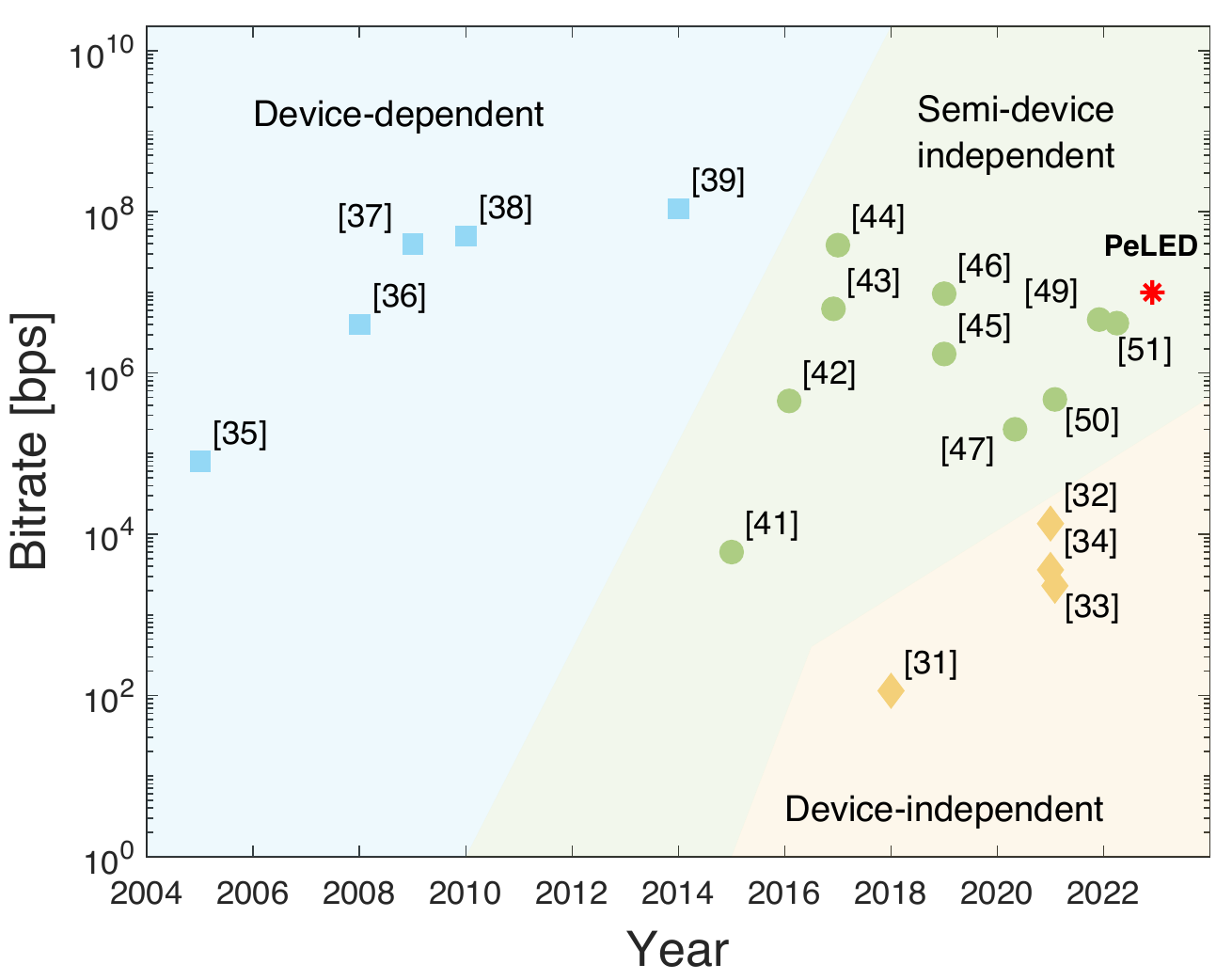}
\caption{\textcolor{black}{Raw random bit generation rate of several quantum random number generators as a function of published year. The results are limited to those that employed single-photon detectors and are based on discrete variable quantum systems for direct comparison purposes. Our maximum sustained raw bit rate is 10.35 Mbit s$^{-1}$.}
\label{Fig1} }
\end{figure}

\textcolor{black}{The highly secure measurement-device independent (MDI) QRNG protocol, which belongs to the SDI class, can remarkably certify private randomness even in the case that the detectors are under the control of an eavesdropper \cite{Supic_2017,Pawlowski_2014}. In the MDI scenario, a user in need of random numbers possesses a characterised state preparation device ($\mathbf{P}$) but has to use an untrusted measurement device ($\mathbf{M}$).} This scenario has become of upmost practical relevance nowadays, as a side-channel attack to the detectors is the most common strategy for hacking quantum technologies \cite{Lo_2012}. 

\section{Measurement-Device-Independent Quantum Random Number Generation using perovskite light emitting diodes}

\subsection{Measurement-device-independent protocol} 
The basic idea is that $\mathbf{P}$ now works actively, by randomly preparing  different  states. One of the states is conventionally used for randomness generation, while the others will be used to check the measurement device. The randomness generation rate is only slightly affected because the testing states can be chosen seldom, and most of the experimental run will still be used for randomness generation. For instance, in our case the state $1/\sqrt{2}(|\textrm{H}\rangle + i|\textrm{V}\rangle)$ is used for randomness generation and the states $|\textrm{H}\rangle$ and $|\textrm{V}\rangle$ will be the testing ones. Whenever the state $|\textrm{H}\rangle$ ($|\textrm{V}\rangle$) is sent, only detector D$_1$ (D$_2$) should fire. Remarkably, from the success probabilities of recording the test states one can put an upper bound on the knowledge an eavesdropper can have about the generated random bit sequence. Formally speaking, and assuming the notation of \cite{Carine_2020}, the preparation device is used to randomly prepare a state of a set, $\{\omega_x\}$, with $x = \{0, 1, 2\}$, which are measured by $\mathbf{M}$, resulting in an outcome $a$. After several iterations, one estimates the probabilities $p(a|\omega_x)$. The maximal probability $P_g(x^*)$ of the eavesdropper guessing the outcome $a$ for a given input $x^*$, compatible with $p(a|\omega_x)$, is given by the solution to a simple numerical optimisation program defined in \cite{Supic_2017, Carine_2020}. Note that the eavesdropper aside of having control of the detectors, is allowed to even resort to quantum entanglement for guessing the outcomes $a$ \cite{Carine_2020}. Finally, the amount of private randomness that is certified per experimental round in the MDI scenario is given by the min-entropy: $\mathcal{H}_{min} (x^*) = -\log_2 P_g(x^*)$.

\subsection{Experimental setup and results}
\textcolor{black}{Our MDI-QRNG is schematically depicted in Fig. \ref{Fig2}a. For each measurement round, the user chooses whether to prepare and send the test states, represented by horizontal or vertical polarisation states ($\omega_0 = |\textrm{H}\rangle$ and $\omega_1 = |\textrm{V}\rangle$), or a linear superposition of both test states [$\omega_2 = |\textrm{R}\rangle = 1/\sqrt{2}(|\textrm{H}\rangle$ +$i|\textrm{V}\rangle$)] as the random number generation state. The states are produced from weak coherent states of light generated by the PeLED, which is housed in the preparation stage. A linear polariser is employed to filter out the unpolarised light produced by the PeLED before passing through a liquid crystal waveplate (LCWP) with a switching time of less than 50 ms, which is the element employed by the user to implement the choice of $\omega_x$. A ball lens with 50 mm focal length is used to collimate the light emitted from the PeLED, and a calibrated optical attenuator is needed to adjust the mean photon number per detection window, such that the probability to have multi-photon events at the single-photon detectors is negligible ($<0.28\%$). Both components are omitted for the sake of simplicity. In the MDI framework, the measurement stage is considered as a black box, which is able to output a signal ``0'' or ``1'' depending on the measurement result. Nothing else needs to be assumed about the measurement procedure to certify the privacy of the generated random numbers. In our implementation, the projection of the prepared states is done with an uncharacterised polarising beam splitter (PBS), which in principle should have the horizontal ($|\textrm{H}\rangle$) and vertical ($|\textrm{V}\rangle$) polarisation states associated to the output signals ``0'' and ``1'', respectively (see Fig. \ref{Fig2}a). The measurement procedure is concluded with silicon single-photon avalanche detectors D$_1$ and D$_2$ (Perkin Elmer), with 25\% overall detection efficiency at 800 nm. The output ``0'' or ``1'' of the measurement stage is registered by field programmable gate array (FPGA) electronics, whose output is then continuously streamed to a personal computer for storage and randomness extraction (see the Appendix). In the event that neither detector records the presence of a photon (for example due to absorption), that detection event is discarded and not stored. Similarly, if both detectors record the presence of a photon (either from noise in the detectors, from photons from the environment that couple to the fibers, or from multiphoton events) that detection event is also discarded.}

\begin{figure}[h!]
\centering
\includegraphics[width=\linewidth]{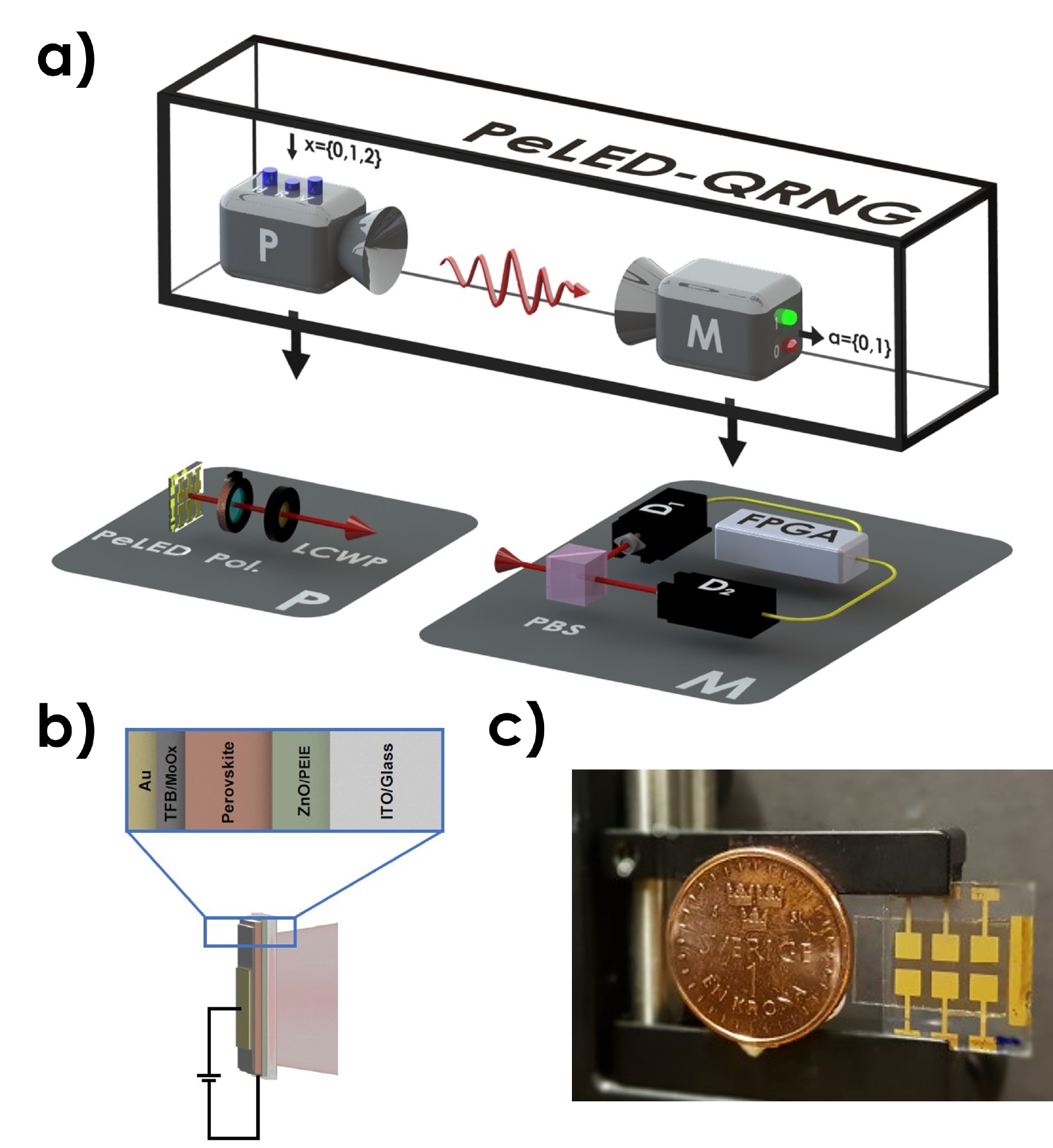}
\caption{\textcolor{black}{Experimental setup and PeLED structure.  a) Scheme of our measurement-device-independent quantum random number generator. On the left the preparation stage $(\mathbf{P})$ creates the quantum states to be prepared depending on the user's choice of test or randomness generation states, while a measurement stage $(\mathbf{M})$ produces the outputs ``0'' or ``1'' for each measurement round. The PeLED is housed inside the preparation stage, where it is used to produce weak coherent states. When the user decides to generate a random number, the superposition state  $|R\rangle = 1/\sqrt{2}(|H\rangle + i|V\rangle)$ is prepared with a fixed linear polariser followed by a liquid crystal waveplate. In order to verify the functioning of the measurement device, the user randomly chooses to prepare one of the test states $|H\rangle$ or $|V\rangle$ by electrically controlling the retardance of the waveplate. In our implementation the measurement device consists of an uncharacterised polarising beam splitter (PBS), which in principle should project the incoming states onto the orthogonal states $|H\rangle$ and $|V\rangle$. The outputs of the PBS are connected to single-photon detectors D$_1$ and D$_2$. The outcomes ``0'' or ``1'' of $(\mathbf{M})$ are read by field programmable gate array (FPGA) electronics.  b) The structure of the employed perovskite diodes, with the biasing circuit indicated. c) Close-up photo of one of the PeLED samples mounted on an opto-mechanical mount. A coin worth one Swedish Crown (SEK) placed as a size reference.}}
\label{Fig2}
\end{figure}

\begin{figure}
\centering\includegraphics[width=\linewidth]{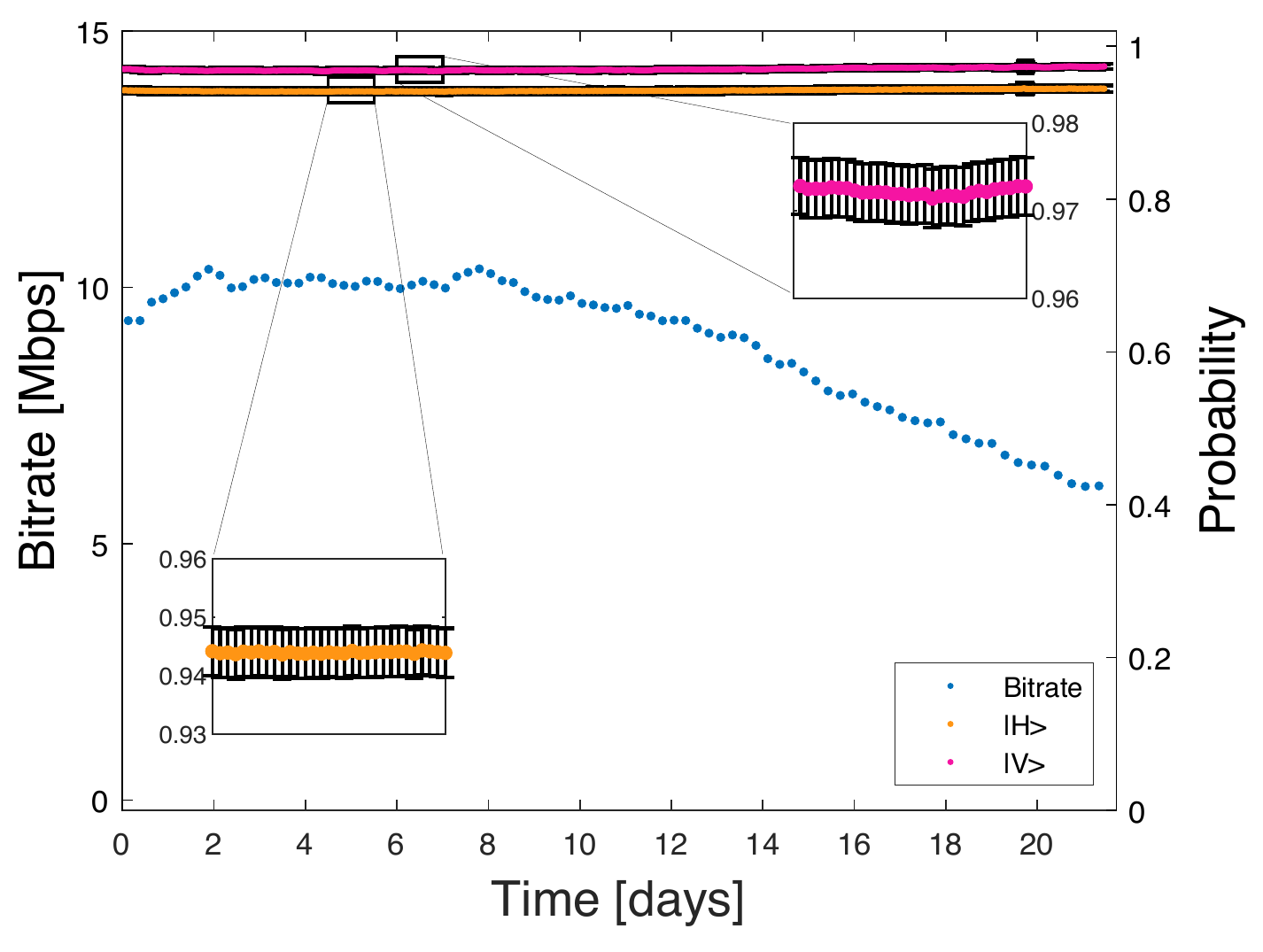}
\centering

\caption{\textcolor{black}{Performance of the quantum random number generator with a PeLED. Random number generation rate as a function of elapsed experiment time. Each blue dot corresponds to the average rate over 6 hours, and we obtain an overall average of $9.01 \pm 1.30$ Mbit s$^{-1}$, with a maximum sustained rate of 10.35 Mbit s$^{-1}$. Also shown are the probabilities to measure either the $|\textrm{H}\rangle$ or $|\textrm{V}\rangle$ test states, as required to test of the measurement device in the MDI protocol. Each test state is randomly chosen by the user with a 0.5\% chance at the beginning of every measurement block. The insets show a higher time resolution sample of each test state data. Error bars come from assuming Poissonian statistics of the PeLED photon number distribution combined with the detection process. }    \label{Fig3}}
\end{figure}

\textcolor{black}{In our experiment, before each measurement data block, a decision is randomly taken on whether that block will be used for randomness generation or to test the measurement device, as required in the MDI QRNG protocol \cite{Carine_2020}. In our experiment, the user choice is provided by an external software-based random number generator, with a bias of 99\% towards randomness generation mode. Please note that any source of randomness may be used, even the partial output of a previous run or another source the user trusts. In Fig. \ref{Fig3} we display the raw bit generation rate as a function of elapsed time of the experiment, where one can see that it attains a remarkably high maximal value of 10.35 Mbit s$^{-1}$ between days 2 and 8 approximately, and an average of $9.01 \pm 1.30$ Mbit s$^{-1}$ over the entire run. We observe stable performance approximately in the first 8 days, and then a slow decay due to degradation of the PeLED, with a constant current density applied of 0.24 mA cm$^{-2}$. Nevertheless after 22 days, the PeLED still has more than half of its original brightness. We also display the success probabilities recorded for either the $|\textrm{H}\rangle$ or $|\textrm{V}\rangle$ test states, when activated by the user, during the whole measurement run. Please note that the user switches seamlessly between the transmitted states with the liquid crystal waveplate. The overall random number generation rate does not suffer a penalty, since the test states are only used 1\% of the time. In spite of the fact the emission rate from the PeLED deteriorates, there is no change in the success probability for the test states, showing there is no change in the certification of the privacy of the random numbers generated. The difference for the measured probabilities between both states, is from the different transmission/reflection coefficients of the polarising beam splitter in the $\mathbf{M}$ stage. The average success probability for the entire generated random bit sequence is $P_{\textrm{suc}}=0.97 \pm 0.01$. This places an upper bound on the randomness generated per experimental round that is certified to be private against detector side-channel attacks. In our particular implementation it corresponds to $\mathcal{H}_\textrm{min}=0.71 \pm 0.01$ random bits, which is quite high considering the innovative aspects of our QRNG.}

\subsection{PeLED Material Structure}
The structure of our PeLED is based on an indium tin oxide (ITO)-coated glass substrate (Fig. \ref{Fig2}b). The active layer consists of formamidinium lead iodide perovskite, with approximately 50 nm thickness. This layer is sandwiched between the electron and hole transport layers which consist of polyethylenimine ethoxylated (PEIE)- modified zinc oxide (ZnO) and poly(9,9-dioctylfluorene-co-N-(4-butylphenyl)diphenylamine) (TFB) ($\sim$ 40 nm). They are followed by layers of molybdenum oxide (MoOx) ($\sim$ 7 nm) and gold ($\sim$ 80 nm) as the contact layer. In order to obtain highly stable PeLEDs, we used an additive of pimelic acid (PAC) \cite{Kuang_2021}. Under forward bias, electrons and holes injected into the perovskite layer recombine, efficiently generating photons with a central wavelength of 804 nm, with a spectral width of 41.6 nm full-width at half-maximum (Supplemental material).

\subsection{Postprocessing and randomness tests}

\begin{figure}
\centering\includegraphics[width=\linewidth]{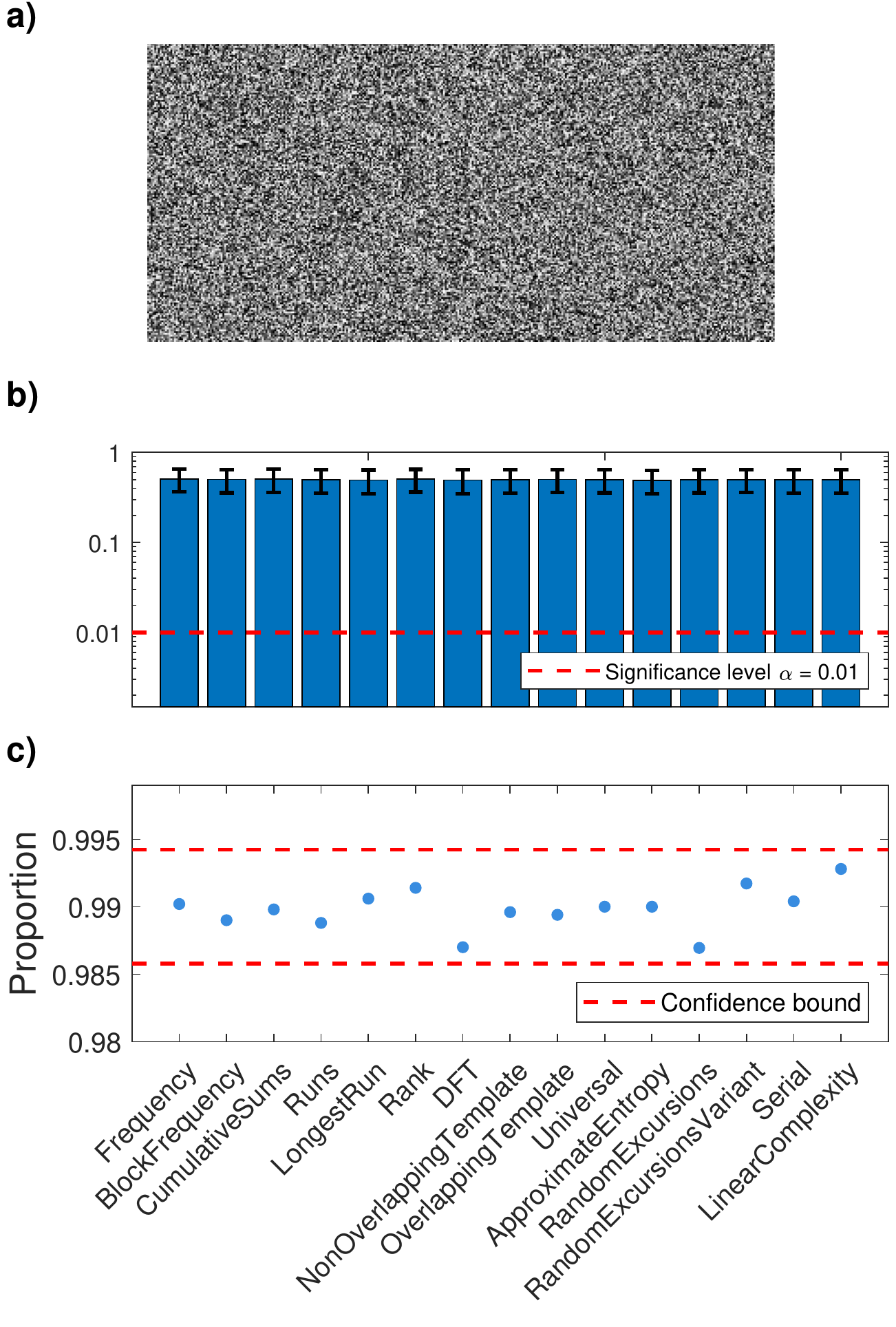}
\centering

\caption{Quantum random number generation with PeLEDs. a) Sequence of generated random bits (0.5 Mbit in length) plotted as an image, where each pixel has an 8-bit gray-scale colour depth. b) Results of the NIST statistical test suite on 5 Gbits of the generated random sequence. Each of the sixteen tests of the suite is labeled on the x-axis, and on the y-axis is the numerical result for each test from 0 to 1, called the p-value. We employ a standard confidence value of 0.01, meaning a test is considered successful if its p-value above this number. Displayed are the average results of each test, with the standard deviation of the test results. c) Here we plot the proportion $r$ of passed runs within each test. The significance level for a test to be considered passed is shown in the graph as the dashed red line (Appendix).     \label{Fig4}}
\end{figure}

Following the Toeplitz randomness extraction procedure we test the randomness of the final random bit key. Initially, we plot the first 0.5 Mbit of the generated sequence as an image where every 8-bit gray-level pixel is encoded directly from the random sequence (Fig. \ref{Fig4}a). We choose only a small subset in this example due to limitations in displaying the entire data as a single image with reasonable size. Nevertheless, one can clearly observe the randomness of the data as the figure is indistinguishable from noise.\textcolor{black}{Then a randomly selected 5 Gbits of the entire generated random bit sequence is fed to the NIST randomness statistical test suite, a well-known benchmark for statistically certifying the randomness of a sequence \cite{NIST}.} The test suite consists of 16 tests (listed in  Fig. \ref{Fig4}b and  Fig. \ref{Fig4}c), and for each test the original sequence is divided into $b$ blocks of equal size. A typical block size is 1 million bits, which we have employed. The test result is a numerical value between 0 and 1 called a p-value, and a test is considered a pass if the corresponding result is above the confidence value of $\alpha = 0.01$. Fig. \ref{Fig4}b shows the average of the p-values for all data blocks in each test with the corresponding standard deviation, clearly showing the generated sequence passes all tests. Furthermore in Fig \ref{Fig4}c we plot the proportion of blocks in each test that passes with a p-value greater than 0.01. Also shown is the confidence level for the minimum proportion of tests that must pass in order ensure randomness certification for the bit sequence (Appendix).

\section{Conclusion and Discussion}

Perovskites have played a major role in the development of highly-efficient solar cells, thus greatly pushing solar power plants into becoming a major player in the production of electricity worldwide. The recent development of optoelectronic devices from perovskites is opening up the field of photonics for this revolutionary material. PeLEDs are poised to have significant impacts in the areas of solid-state lighting and optical communications, with its lower carbon footprint during its entire life cycle as well as optical and mechanical properties being major motivators for their use. These properties are extremely attractive to the blooming field of quantum technologies, specifically aiming towards the widespread deployment of applications, even down to the consumer level. Following this direction, we have showed for the first time that PeLEDs can be successfully employed to implement modern and highly secure quantum random number generation protocols. Our results (Fig. \ref{Fig1}) place our QRNG among the best reported, including commercial devices \cite{Idquantique}, and thus allow the kick-start of a new multi-disciplinary field concerning quantum technology systems with perovskite light sources, while also pointing towards initiatives to make the quantum technology industry more sustainable. We successfully demonstrated that the generated sequence passes the stringent NIST randomness test suite, a widely recognized benchmark employed for the certification of randomness in a sequence of numbers. At the same time, we have also guaranteed the privacy of at least 71\% of the generated random numbers by implementing our QRNG with the MDI approach. These results cement the start of a new line of research into perovskite quantum light sources aimed at producing more compact, robust and\textcolor{black}{sustainable} quantum technology devices that can eventually make its way onto consumer products, thus helping the democratisation of access to quantum technologies.



\section*{Acknowledgments}
We thank D. Cavalcanti and G. Ca\~nas for helpful discussions. We acknowledge CENIIT Link\"{o}ping University, the Swedish Research Council (VR 2017-04470), QuantERA grant SECRET (VR 2019-00392), the Knut and Alice Wallenberg Foundation through the Wallenberg Center for Quantum Technology (WACQT), ERC Starting Grant (No. 717026) and the Wallenberg Academy Fellowship for financial support. G.L. was supported by Fondo Nacional de Desarrollo Cient\'{i}fico y Tecnol\'{o}gico (FONDECYT) (1200859) and ANID - Millennium Science Initiative Program - ICN17\_012).

\section*{Appendix}

\subsection{Randomness extractor and data streaming}

\textcolor{black}{The generated raw bit sequence is buffered in the FPGA in blocks of $2^{16}$ bits before they are streamed over a UDP protocol to a host computer acting as server for final storage. The server then stores the acquired data for each of the different states separately to facilitate i) security analysis, and ii) randomness extraction.} We then generate close to perfectly uniform sequences of numbers by applying a Toeplitz hashing extractor to the generated raw bits recorded for the experiment rounds where the prepared state was $\ket{\textrm{R}}$. We first split the raw binary sequence into $N$ sequences of length $n$ which we then multiply with an $n \times m$ Toeplitz matrix to generate an output of $m$ hashed bits, following the procedure in \cite{Qi_2017}. Therefore we employ $n = 400$, and $m$ is given by $m=\mathcal{H}-2\log_2 \varepsilon$ where $\mathcal{H} = -\log_2\left[ \max_{x\in \left\{0,1\right\}^8} \mathrm{Pr}\left\{X=x\right\} \right]$ is the minimum entropy when considering binary strings of length 8, and $\varepsilon=2^{-100}$ is a security parameter derived from the leftover hash lemma \cite{Ma2013}.

The Toeplitz matrix $T$ is defined at the beginning of the extraction procedure using a seed of $n+m-1$ bits taken from the beginning of the raw sequence. These bits specify the elements of the first row and column of the matrix. As $T$ can be reused for each of the subsequences, we obtain the final extracted sequence by concatenation of the results of the matrix multiplications $Tn_1, Tn_2, ..., Tn_N$, where the subscript indicates the sequence number \cite{Ma2013, Qi_2017}.

\subsection{Statistical testing}
For each of the tests in the NIST 800-22 suite we compute the proportion $r$ of sequences that yielded a p-value greater than the significance level $\alpha = 0.01$ \cite{NIST}. We employ a confidence interval for the proportion of passed tests as $r \pm 3\sqrt{\frac{\alpha(1-\alpha)}{b}}$, as defined in \cite{NIST}, where $b=5000$ is the number of sequences fed to the test. This results in a confidence interval of $0.9858 \leq r \leq 0.9942$, where we consider a test passed if the proportions of sequences with p-values $p_i\geq \alpha$ fall within the interval.

\section*{Supplemental Material}

\subsection*{Fabrication and characterisation of the PeLEDs}

The detailed procedure of PeLEDs fabrication can be found in our previous study \cite{Kuang_2021}. Briefly the cleaned ITO glass was spin coated with the ZnO layer under ambient conditions and then transferred to a nitrogen-filled glovebox where the PEIE, formamidinium lead iodide perovskite (formamidinium iodide (FAI):PbI$_2$:PAC with molar ratios of 3:1:0.5 in dimethylformamide as the precursor solution) and TFB layers were spin coated and annealed (if necessary) in sequence. The MoOx and Au layers were evaporated and deposited at a base chamber pressure of $\sim 2 \times 10^{-6}$ mbar with a shadow mask to define the device area as 0.0725 cm$^2$. 

Electroluminescence performance was measured in the nitrogen filled glovebox. The optical spectrum was measured by a fibre integration sphere (FOIS-1) coupled with a QE Pro spectrometer (Ocean Optics), and is plotted in Fig. \ref{Fig1}. The centre wavelength is at 804 nm, with a spectral width of 41.6 nm (full-width at half-maximum). 

\begin{figure}
\centering\includegraphics[width=10cm]{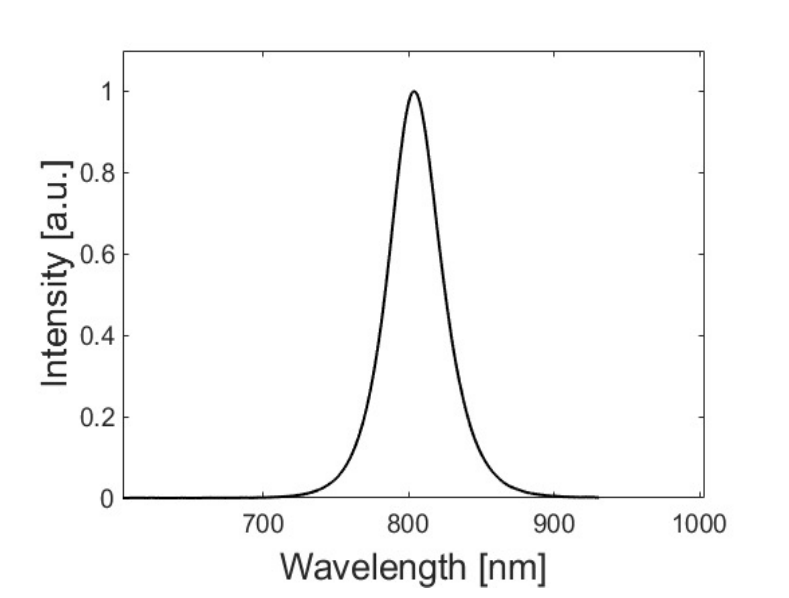}
\caption{Emission spectrum of the PeLED. \label{Fig1}}
\end{figure}

A Keithley 2400 sourcemeter was then used to apply a voltage bias to the device and simultaneously record the current flowing through it. During the test, only light emitting from the glass side of the device was collected. A home-made Labview program was used to control the test, collect data, and calculate the performance metrics (EQE, Radiance, etc...) based on the collected data. EQE is a parameter that describes the efficiency of an LED which is defined by the ratio of the number of emitting photons to the number of injected carriers. The EQE can be obtained from the collected spectrum and current density described by the following equation:

\begin{equation}
    EQE = \frac{\int\frac{\phi(\lambda)}{h\frac{c}{\lambda}}d\lambda}{\frac{JS}{q}}
\end{equation}

where $\phi(\lambda)$ is the spectral intensity (with units of W$\cdot$ nm$^{-1}$), with wavelength $\lambda$, $h$ is Planck's constant, $c$ the speed of light, $J$ is the measured current density, $S$ the device area of the LED and $q$ is the elementary charge. Radiance (with units of W$\cdot$m$^2 \cdot$Sr$^{-1}$) is a metric to the emitted light power at a given solid angle of an LED, which is obtained by integrating the collected spectrum:

\begin{equation}
R = \frac{\int \phi(\lambda)d\lambda}{\pi S}
\end{equation}

The current density and the radiance for a given applied voltage is given in Fig. \ref{Fig2}a, while the external quantum efficiency (EQE) is plotted in Fig. \ref{Fig2}b as a function of the current density. More details of the fabrication and characterisation of the PeLEDs can be found in \cite{Xu_2019, Kuang_2021}.

\begin{figure*}[h!]
    \centering\includegraphics[width=15cm]{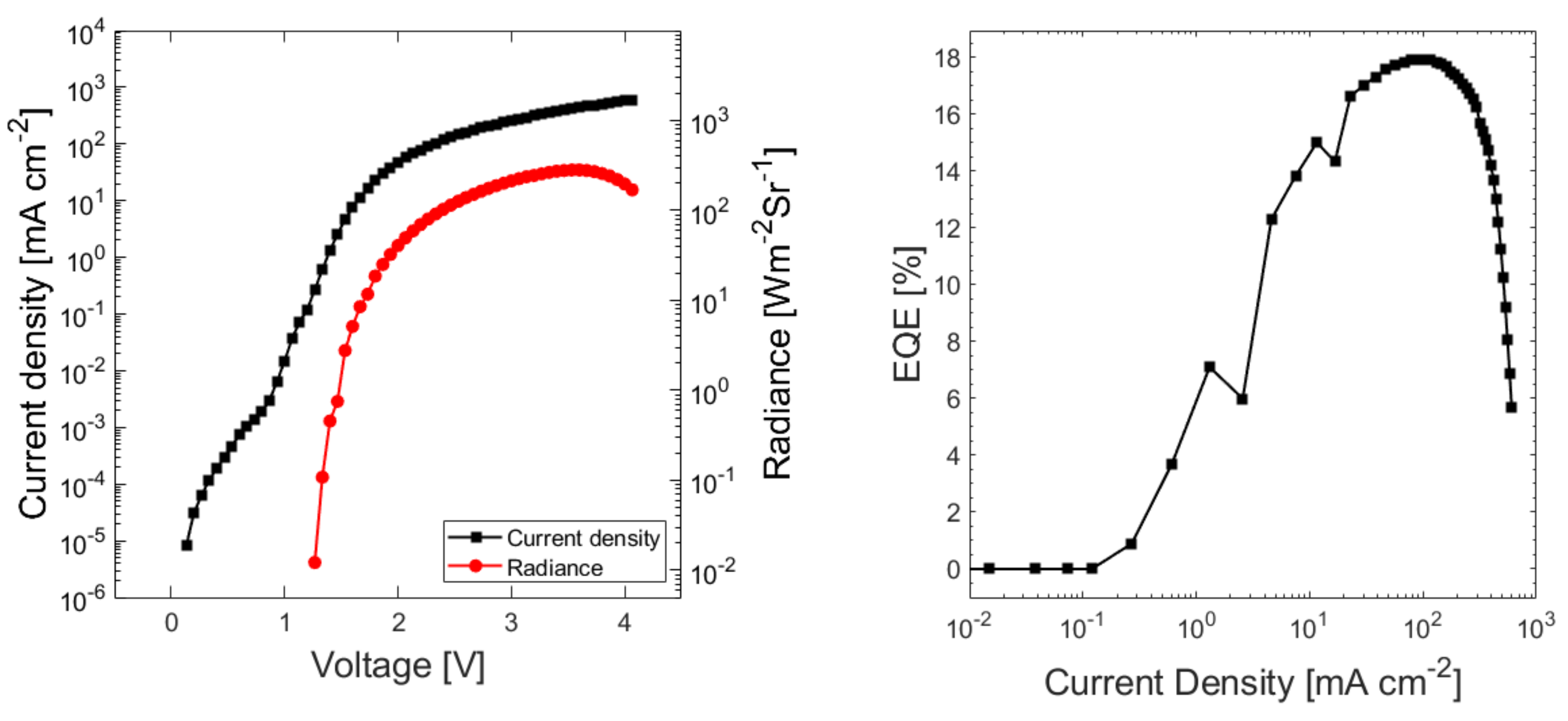}
    \caption{Electro-optical characteristics of the employed PeLEDs. a) Current density and the radiance emitted by the PeLED as a function of the applied voltage. b) External quantum efficiency as a function of the current density.\label{Fig2}}
    \end{figure*}

\subsection*{Measured photon number distribution}

\begin{figure*}[h!]
    \centering
    \includegraphics[width=12cm]{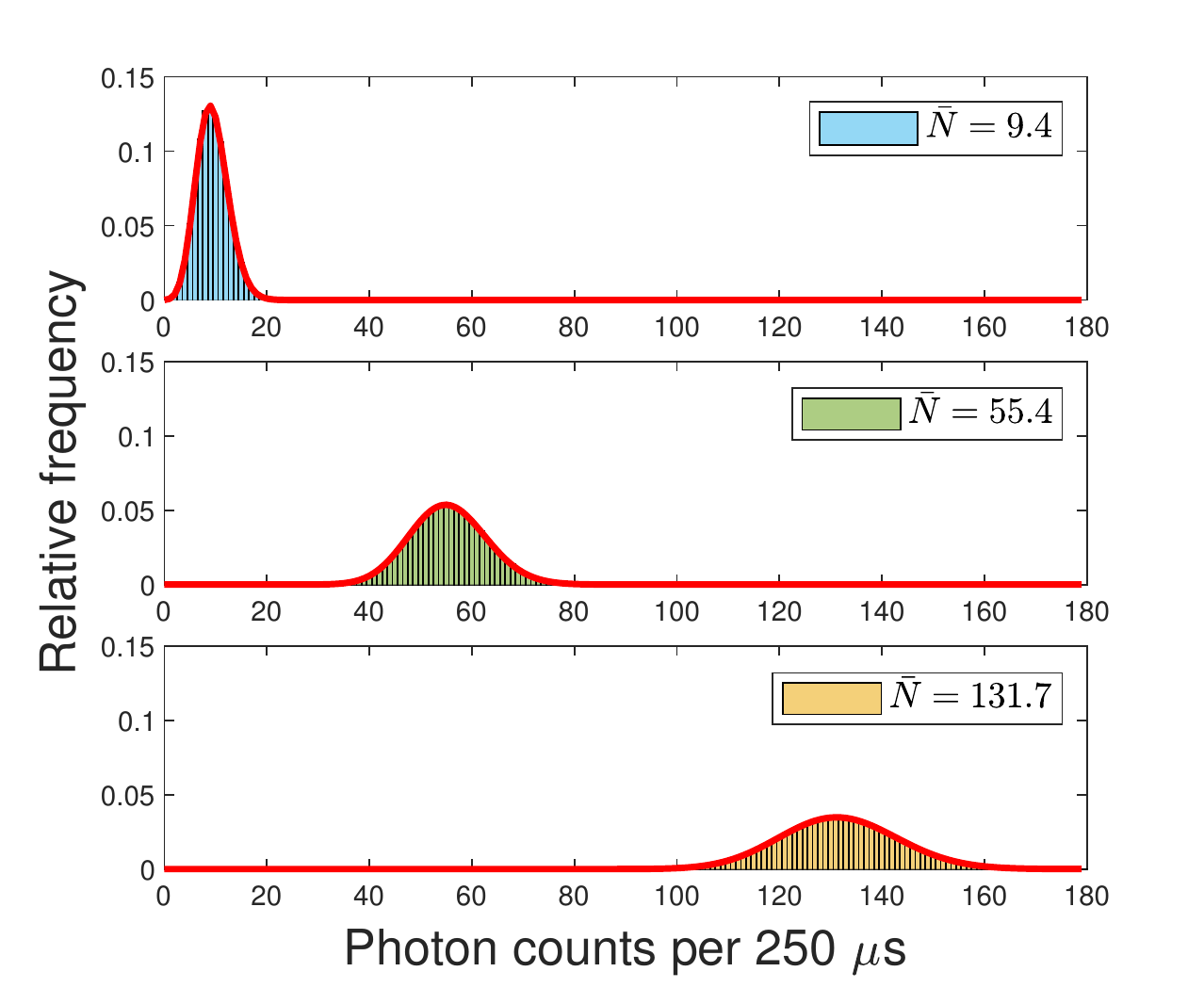}
    \caption{\label{fig:poisson}Characterization of the photon number distribution for different mean photon numbers $\bar{N}$ and the corresponding theoretical Poisson photon number distribution (red line). The histogram shows the recorded photon counts $n$ per $\tau = 250 \mu s$ integration time.}
\end{figure*}

In order to characterise the photon statistics of our device combined with the detection system, we record the photon count number $n$ during a time window $\tau$ of 250 $\mu$s emitted directly from the device, for three different values of the mean photon number $\bar{N}$, adjusted with the electric current directly biasing the PeLED. The experiment is run several times for each $\tau$ interval, and the histogram of obtained $n$ counts is plotted in Fig \ref*{fig:poisson}. We fit in each case the poissonian probability distribution $P_n = (\bar{N}^ne^{-\bar{N}})/n!$. As can be seen the recorded photon number distribution  follows very closely the expected prediction.


\end{document}